\documentstyle[pra,aps,epsf]{revtex}

\newcommand{\cm}{\,\mathrm{cm}}
\newcommand{\eV}{\,\mathrm{eV}}
\newcommand{\Ry}{\,\mathrm{Ry}}
\newcommand{\icm}{{\,\mathrm{cm^{-1}}}}
\newcommand{\dif}{\,\mathrm{d}}
\newcommand{\ldirac}[1]{\langle #1 |}
\newcommand{\rdirac}[1]{|#1\rangle}
\renewcommand{\vec}[1]{\mbox{\boldmath$#1$}}

\draft

\begin{document}

\title{Enhancement of the electron electric dipole moment in gadolinium 3+}
\author{S. Y. Buhmann, V. A. Dzuba and O.P. Sushkov}

\address{School of Physics, University of New South Wales,\\
 Sydney 2052, Australia}

\maketitle

\begin{abstract}

There have been recent suggestions for searching for the electron
electric dipole moment, using solid state experiments with
compounds containing Gd$^{3+}$ ions \cite{Lam,Hun}. These
experiments could improve the sensitivity compared to present
atomic and molecular experiments by several orders of magnitude.
The analysis of the problem requires a calculation of the
enhancement coefficient $K$ for the electron electric dipole
moment in the Gd$^{3+}$ ion. In this work we perform this
calculation. The result is $K \approx -4.9\pm 1.6$. Limitations of
the accuracy of the calculation are mainly due to the lack of data on
Gd$^{3+}$ excitation spectra. We formulate which quantities have
to be measured and/or calculated to improve the accuracy.

\end{abstract}

\pacs{PACS: 11.30.Er, 32.10.Dk, 31.25.Eb}

\section{introduction}

A violation of the combined symmetry of charge conjugation (C) and
parity (P) has been discovered in the decay of the $K^0$ meson
about 40 years ago \cite{CCK}. The exact origin of this symmetry
violation remains an enigma, although the standard model of
electroweak interactions can describe these processes
phenomenologically. It has also been proposed by Sakharov
\cite{Sak} that the matter-antimatter asymmetry observed in our
universe could have arisen from a CP-violating interaction active
at an early stage of the big bang. The CP-violation implies a
time-reversal (T) asymmetry and hence violation of the combined
TP-symmetry, because there are strong reasons to  believe that the
combined CPT-symmetry should not be violated \cite{CPT}. An
electric dipole moment (EDM) of a system in a stationary quantum
state indicates a violation of the TP-symmetry; this is why
searches for EDMs of elementary particles, atoms and molecules are
a very important approach to the studies of violations of
fundamental symmetries \cite{KL}. In the present work we
concentrate on the EDM of the electron $d_e$.

At present the best limitation on $d_e$ comes from the Berkeley
experiment with an atomic thallium beam \cite{Com}, $d_e <
1.6\cdot 10^{-27}\, e\cm$. There are some ideas for improving the
sensitivity. One way of improvement is an experiment with
metastable levels of PbO molecules \cite{DeM}. A breakthrough
could be achieved in solid state experiments. This idea was
suggested by Shapiro in 1968 \cite{Sh}. The application of strong
electric fields to electrons bound within a solid would align the
electric dipole moments of these electrons. This should lead to a
simultaneous alignment of the electron spins; the magnetic field
arising from this alignment could be detected experimentally. An
experiment of this kind has been performed with nickel-zinc
ferrite \cite{VK}. However, due to experimental limitations, the
result was not very impressive. Interest to the approach has been
renewed recently due to suggestions of Lamoreaux \cite{Lam} and
Hunter \cite{Hun} to perform similar experiments with gadolinium
gallium garnet, Gd$_3$Ga$_5$O$_{12}$, and gadolinium iron garnet,
Gd$_3$Fe$_5$O$_{12}$, employing new experimental techniques. First
estimates of sensitivity promise to improve the current upper
limit on the electron EDM by at least three orders of magnitude,
depending on the experimental setup. A thorough analysis of the
problem requires the calculation of the enhancement coefficient
for the electron EDM in the Gd$^{3+}$ ion. We perform this
calculation in the present work.

\section{Single particle contribution}

The Gd$^{3+}$ ion has a nucleus with charge $Z=64$ and 61
electrons. The 7 electrons in the outer shell occupy the $4f$
orbitals. So the shell is half filled, and hence the total orbital
angular momentum is zero, $L=0$, and total spin $S=7/2$. Let us
consider the state with maximum z-projection of the spin. In the
representation of second quantization the ground state wave
function is of the form
\begin {equation}
\label{0}
|gs\rangle=f_{-3\uparrow}^{\dag}f_{-2\uparrow}^{\dag}f_{-1\uparrow}^{\dag}
f_{0\uparrow}^{\dag}f_{1\uparrow}^{\dag}f_{2\uparrow}^{\dag}
f_{3\uparrow}^{\dag}|0\rangle,
\end{equation}
where $f_{m\sigma}^{\dag}$ is the creation operator for a $4f$
electron with spin $\sigma$ and  z-projection of orbital angular
momentum $l_z=m$. The TP-odd interaction of the electron EDM with
the electric field $\vec{E}$ is of the form, see e.g. Ref.
\cite{KL}
\begin{equation}\label{Vd}
 V_d=-d_e\gamma_0\vec{\Sigma\cdot E},
\end{equation}
where $\gamma_0$ and $\vec{\Sigma}=\gamma_0\gamma_5 \vec{\gamma}$
are Dirac $\gamma$-matrices. Because of Schiff's theorem
\cite{Sch} it is crucially important to account for very complex
many-body screening effects, when working with the Hamiltonian
(\ref{Vd}); technically this means that the many-body perturbation
theory practically is not convergent. The standard way \cite{KL}
to avoid this complication is to split the Hamiltonian into two
terms: $V_d=-d_e\gamma_0\vec{\Sigma\cdot E}=-d_e\vec{\Sigma\cdot
E} -d_e(\gamma_0-1)\vec{\Sigma\cdot E}$. Then due to Schiff's
theorem \cite{Sch} the contribution of the first term to the EDM
of the ion is identically zero, so one can reduce the interaction
\begin{equation} \label{Vd1}
V_d \to V_d^r=-d_e(\gamma_0-1)\vec{\Sigma\cdot E}.
\end{equation}
Perturbation theory with this operator is reasonably convergent.

\begin{figure}[h]
\vspace{2pt} \hspace{-35pt} \epsfxsize=8cm
\centering\leavevmode\epsfbox{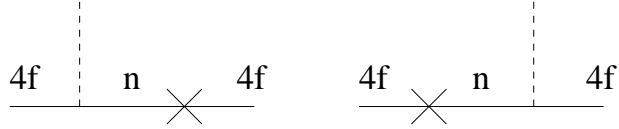} \vspace{8pt} \caption{\it
{Leading contribution to the Gd$^{3+}$ EDM. The dashed line
denotes the dipole moment $ez=er\cos\theta$, and the cross denotes
the TP-odd reduced interaction $V_d^r$, eq (\ref{Vd1}).}}
\label{Fig1}
\end{figure}

In leading order of single particle perturbation theory the EDM of
the ion $D_{\mathrm{sp}}$ is given by diagrams shown in
Fig.\ref{Fig1}.
\noindent The corresponding formula reads
\begin{equation} \label{D}
 D_{\mathrm{sp}}=2\sum_{mn}\frac{\ldirac{4f_{m\uparrow}}er\cos\theta\rdirac{n}
 \ldirac{n}V_d^r\rdirac{4f_{m\uparrow}}}{E_{4f}-E_n}=K_{\mathrm{sp}} d_e,
\end{equation}
where the EDM enhancement coefficient is given by
\begin{equation} \label{K}
 K_{\mathrm{sp}}=2\sum_{mn}\frac{\ldirac{4f_{m\uparrow}}(r/a_B)\cos\theta\rdirac{n}
 \ldirac{n}(e a_B/d_e)V_d^r\rdirac{4f_{m\uparrow}}}{E_{4f}-E_n}.
\end{equation}
Here $e$ is the electron charge and $a_B$ is the Bohr radius. We
first consider $5d$ and $5g$ intermediate states, as the
contribution of other states is much less. The E1 matrix elements
are of the form
\begin{eqnarray} \label{E1}
 \ldirac{4f_{m\uparrow}}(r/a_B)\cos\theta\rdirac{5d_{m\uparrow}}
&=&\sqrt{\frac{9-m^2}{35}}\,r_{4f,5d},\\
\ldirac{4f_{m\uparrow}}(r/a_B)\cos\theta\rdirac{5g_{m\uparrow}}
&=&\sqrt{\frac{16-m^2}{63}}\,r_{4f,5g}, \nonumber
\end{eqnarray}
where  $r_{ik}$ is the E1-transition radial integral expressed in
atomic units (Bohr radius),
\begin{equation} \label{rik}
 r_{ik}= \int_0^{\infty}R_i(r)R_k(r)r^3\dif r.
\end{equation}
$R_i(r)$ is the radial wave function of the corresponding orbital,
$\int_0^\infty R_i^2(r)r^2dr = 1$.

The matrix element of $V_d^r$ has been calculated earlier, see
e.g. Ref.\cite{KL}. In jj coupling scheme, and using a
semiclassical approximation, it reads
\begin{equation} \label{VV}
 \ldirac{n'j\,l\pm
1}(ea_B/d_e)V_d^r\rdirac{nj\,l}=-\frac{4Z^3\alpha^2}{\gamma(4\gamma^2-1)}
\cdot\frac{Z_{\mathrm{eff}}^2}{(\nu'\nu)^{3/2}}\,E_0,
\end{equation}
where $\alpha=1/137.036$ is the fine structure constant,
$\gamma=\sqrt{(j+1/2)^2-(Z\alpha)^2}$, $E_0=2\Ry =27.2\eV =2\times
109,737\icm$ is the atomic energy unit, $Z_{\mathrm{eff}}=4$ is
the effective charge for electron motion at large distances, and
finally the effective principal quantum numbers $\nu$ and $\nu'$
are defined by the electron energy with respect to the ionization
limit, $E=-Z_{\mathrm{eff}}^2 E_0 /(2\nu^2)$. Our numerical
calculations in Hartree-Fock approximation agree with the
semiclassical result (\ref{VV}) within ten per cent. The ground
state wave function (\ref{0}) is given in LS coupling scheme,
therefore we have to rewrite the operator $V_d^r$ in this
representation. A simple calculation gives
\begin{eqnarray} \label{LS}
 \ldirac{5d_{m\uparrow}}(ea_B/d_e)V_d^r\rdirac{4f_{m\uparrow}}&=&
-\sqrt{\frac{9-m^2}{35}}\,\ldirac{5d_{5/2}}(ea_B/d_e)V_d^r\rdirac{4f_{5/2}},\\
\ldirac{5g_{m\uparrow}}(ea_B/d_e)V_d^r\rdirac{4f_{m\uparrow}} &=&
-\sqrt{\frac{16-m^2}{35}}\,\ldirac{5g_{7/2}}(ea_B/d_e)V_d^r\rdirac{4f_{7/2}}.\nonumber
\end{eqnarray}
Combining equations (\ref{K}),(\ref{E1}),(\ref{VV}), and
(\ref{LS}) and performing the summation over $m$, we find the
following expression for the single particle contribution to the
enhancement coefficient
\begin{equation} \label{KSP}
 K_{\mathrm{sp}}=8Z^3\alpha^2Z_{\mathrm{eff}}^2\left(\frac{r_{4f,5d}}
{\gamma_{5/2}(4\gamma_{5/2}^2-1)(\nu_{4f}\nu_{5d})^{3/2}}\cdot
\frac{1}{(E_{4f}-E_{5d})}+\frac{4}{3}\cdot\frac{r_{4f,5g}}
{\gamma_{7/2}(4\gamma_{7/2}^2-1)(\nu_{4f}\nu_{5g})^{3/2}}\cdot
\frac{1}{(E_{4f}-E_{5g})}\right).
\end{equation}
Note that from here on we skip the atomic energy unit $E_0$ in all
equations, assuming that all energies are expressed in units of
$E_0$. To find values of the parameters that appear in this
equation, we have performed a Hartree-Fock calculation for the
Gd$^{3+}$ ion. It is known that a Hartree-Fock calculation for an
open shell is not a uniquely defined procedure. In our calculation
we used the following averaging: we assumed that the shell is
fully occupied, but the occupation number of each single particle
$f$-orbital is 1/2. This is a crude approximation, it gives
reasonable values of energy levels and radial integrals, but one
cannot rely on this calculation as far as energy splittings are
concerned. As a result of the calculation we obtained the
following values of energies and radial integrals:
\begin{eqnarray}
\label{hf} &&E_{4f}=-1.65 \ \ \ \ E_{5d}=-1.20 \ \ \ \
E_{5g}=-0.32\\ &&r_{4f,5d}=0.63 \ \ \ \ r_{4f,5g}=0.088\nonumber
\end{eqnarray}
Where available, experimental data should be used, but
unfortunately the experimental data on this particular ion is
scarce. Only the value of the $4f$ energy level (ionization limit)
is known, see Ref. \cite{EL},
\begin{equation} \label{exp}
 E_{4f}=-355000\icm=-1.62\,E_0.
\end{equation}
It agrees well with (\ref{hf}). From (\ref{exp}) and (\ref{hf})
one finds $\nu_{4f}=2.22$, $\nu_{5d}=2.58$ and $\nu_{5g}=5.0$.

The most important is the $E_{4f}-E_{5d}$ energy splitting. The
accuracy of the present Hartree-Fock calculation is not sufficient
to determine this splitting. There is reliable experimental data
for the energy levels of Gd$^{2+}$, Gd$^{1+}$, and Eu$^{2+}$, see
Ref.\cite{EL}. Naive extrapolation of the splitting from these
ions gives
\begin{equation}
\label{A}
A: \ \ \ \  E_{5d}-E_{4f}\approx 40,000\icm \approx 0.18 \,E_0.
\end{equation}
On the other hand there is experimental data for Gd$^{3+}$
\cite{Gde} that indicates
\begin{equation}
\label{B} B: \ \ \ \ E_{5d}-E_{4f}\approx 100,000\icm \approx 0.45
\,E_0.
\end{equation}
Unfortunately Ref.\cite{Gde} does not contain identifications of
all the possible levels, therefore we cannot quite rely on the
data. For this reason we will present two estimates of $K$: one
for the case (A), see (\ref{A}), and another for the case (B), see
(\ref{B}). Experimental and/or theoretical determination of the
$E_{5d}-E_{4f}$ splitting would be the most important to improve
the accuracy of the calculation of the enhancement coefficient
$K$.

Substituting values of the parameters listed above into eq.
(\ref{KSP}) we find the contribution of the $5d$ intermediate
state to the EDM enhancement coefficient:
\begin{eqnarray}
\label{KSP1}
&&A: \ \ \ \ K_{\mathrm{5d}}=-4.5,\\
&&B: \ \ \ \ K_{\mathrm{5d}}=-1.8\nonumber
\end{eqnarray}
The contribution of the $5g$ intermediate state is very small. We
also estimate the contribution of the higher $d$-levels (mainly
the continuous spectrum) as
\begin{equation}
\label{KSP2} A,B: \ \ \ \ K_{{\mathrm{nd,n>5}}} \approx -1.
\end{equation}
Altogether this gives the following value of the single particle contribution
to the EDM enhancement coefficient
\begin{eqnarray}
\label{KSP3}
&&A: \ \ \ \ K_{\mathrm{sp}}=-5.5,\\
&&B: \ \ \ \ K_{\mathrm{sp}}=-2.8\nonumber
\end{eqnarray}

\section{Many-body corrections}
\begin{figure}[h]
\vspace{2pt} \hspace{-35pt} \epsfxsize=10cm
\centering\leavevmode\epsfbox{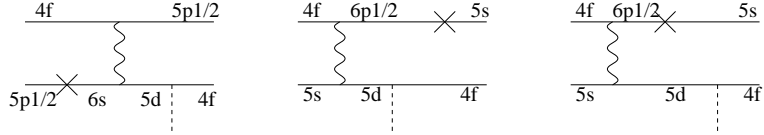} \vspace{8pt} \caption{\it
{Leading many-body corrections to the Gd$^{3+}$ EDM. The dashed
line denotes the dipole moment $er\cos\theta$, the cross denotes
the TP-odd reduced interaction $V_d^r$, eq (\ref{Vd1}), and the
wavy line denotes residual Coulomb interaction}} \label{Fig2}
\end{figure}

In the situation with cesium or any other atom with valent $s$- or
$p$-electrons \cite{KL} the single particle estimate is
satisfactory. However here we have $f$-electrons that have a very
small wave function in the vicinity of the nucleus. As a result
the single particle contribution is strongly suppressed.
Technically this suppression is reflected in eq. (\ref{VV}); the
matrix element is proportional to $1/\gamma^3$, and for
$f_{5/2}$-electrons $\gamma \approx 3$, while for $s_{1/2}$- or
$p_{1/2}$-electrons $\gamma\approx 0.88$. Therefore it is very
important to estimate the many-body corrections. The leading
many-body corrections  to the EDM enhancement coefficient
$K_{\mathrm{mb}}$ are shown in diagrams Fig.\ref{Fig2}, where the
wavy line denotes residual Coulomb interaction
\begin{equation} \label{VC}
V_C=\frac{1}{|\vec{r_i}-\vec{r_j}|}
=\sum_{k=0}^\infty\sum_{q=-k}^{k}
\frac{4\pi}{2k+1}\cdot\frac{r_<^k}{r_>^{k+1}}Y_{kq}^\ast(\Omega_i)Y_{kq}(\Omega_j).
\end{equation}
We only account for diagrams with momentum of the Coulomb quantum
$k$ not higher than 2. The contribution of each diagram from
Fig.\ref{Fig2} must be doubled, because an opposite order of
operators is also possible.
\noindent The formula for the many-body contribution to the
enhancement coefficient due to the diagrams in Fig.\ref{Fig2}
reads
\begin{eqnarray} \label{KMB}
 K_{\mathrm{mb}}=-2\sum_m&&\left(\frac{
\ldirac{4f_m}(r/a_B)\cos\theta\rdirac{5d_m}
\ldirac{5p_{1/2}}\ldirac{5d_m}V_C\rdirac{6s}\rdirac{4f_m}
\ldirac{6s}(ea_B/d_e)V_d^r\rdirac{5p_{1/2}} }
{\left(E_{4f}-E_{5d}\right)\left(E_{5p}-E_{6s}\right)}\right.\\
\nonumber &&+\frac{ \ldirac{5s}(ea_B/d_e)V_d^r\rdirac{6p_{1/2}}
\ldirac{4f_m}(r/a_B)\cos\theta\rdirac{5d_m}
\ldirac{6p_{1/2}}\ldirac{5d_m}V_C\rdirac{5s}\rdirac{4f_m}}
{\left(E_{5s}-E_{6p}\right)\left(E_{4f}+E_{5s}-E_{5d}-E_{6p}
\right)}\\ \nonumber
 &&+\left.\frac{
\ldirac{4f_m}(r/a_B)\cos\theta\rdirac{5d_m}
\ldirac{5s}(ea_B/d_e)V_d^r\rdirac{6p_{1/2}}
\ldirac{6p_{1/2}}\ldirac{5d_m}V_C \rdirac{5s}\rdirac{4f_m}}
{\left(E_{4f}-E_{5d}\right)\left(E_{4f}+E_{5s}-E_{5d}-E_{6p}\right)}\right).
\nonumber
\end{eqnarray}
We have only included $s-p_{1/2}$ matrix elements of $V_d^r$ and
only the intermediate states involving $5d$ electrons which give
the main contribution. All the diagrams in Fig.\ref{Fig2} are
exchange ones, this is why the sign in eq.(\ref{KMB}) is negative.

Matrix elements of the Coulomb interactions are of the form
\begin{eqnarray} \label{E2}
\ldirac{5p_{1/2}}\ldirac{5d_m}V_C\rdirac{6s}\rdirac{4f_m}
&=&-\frac{1}{5}\sqrt{\frac{9-m^2}{35}}\,F^{(2)}(5p,4f;5d,6s)\\
\nonumber
\ldirac{6p_{1/2}}\ldirac{5d_m}V_C\rdirac{5s}\rdirac{4f_m}
&=&-\frac{1}{5}\sqrt{\frac{9-m^2}{35}}\,F^{(2)}(6p,4f;5d,5s)
\end{eqnarray}
where $F^{(k)}(i,j;h,l)$ is the usual Coulomb  radial integral,
\begin{equation} \label{rijkl}
F^{(k)}(i,j;h,l)=\int_0^{\infty}\int_0^{\infty}\frac{r_<^k}{r_>^{k+1}}
R_i(r_1)R_j(r_1)R_h(r_2)R_l(r_2) r_1^2r_2^2\dif r_1\dif r_2,
\end{equation}
expressed in atomic units.

Using matrix elements (\ref{E1}), (\ref{VV}) and (\ref{E2}), and
performing the summation over $m$ in (\ref{KMB}), we obtain the
following expression for the leading many-body correction to the
enhancement coefficient
\begin{eqnarray} \label{KMB1}
K_{\mathrm{mb}}
 &=&-{{8Z^3\alpha^2Z_{\mathrm{eff}}^2 }\over{
5\gamma_{1/2}(4\gamma_{1/2}^2-1)}}r_{4f,5d}
\left\{{{F^{(2)}(5p,4f;5d,6s)}\over{(\nu_{6s}\nu_{5p})^{3/2}}}
\cdot\frac{1}{(E_{4f}-E_{5d})}\cdot\frac{1}{(E_{5p}-E_{6s})}\right.\\
\nonumber
 &&+\left.
{{F^{(2)}(6p,4f;5d,5s)}\over{(\nu_{5s}\nu_{6p})^{3/2}}}
\cdot\frac{1}{(E_{4f}-E_{5d}+E_{5s}-E_{6p})}
\left(\frac{1}{(E_{4f}-E_{5d})}+\frac{1}{(E_{5s}-E_{6p})}\right)
\right\}
\end{eqnarray}
Like the parameters presented in (\ref{hf}) the energy levels and
radial integrals in this formula have been calculated using a
Hartree-Fock method
\begin{eqnarray} \label{hf1}
&&E_{5s}=-3.15 \ \ \ \ E_{5p}=-2.26 \ \ \ \ E_{6s}=-1.03 \ \ \ \
E_{6p}=-0.84 \\ \nonumber
 &&F^{(2)}(5p,4f;5d,6s)=0.028 \ \ \ \ F^{(2)}(6p,4f;5d,5s)=-0.023.
\end{eqnarray}
The corresponding effective principal quantum numbers are
$\nu_{5s}=1.59$, $\nu_{5p}=1.88$, $\nu_{6s}=2.78$,
$\nu_{6p}=3.09$. Using the $E_{5d} - E_{4f}$ energy splitting
given in (\ref{A}) and (\ref{B}) and substituting all the
parameters into eq. (\ref{KMB1}), we find the many-body correction
due to the $6s$-, $6p$-, and $5d$- intermediate states, see Fig.2
\begin{eqnarray}
\label{KMB2}
&&A: \ \ \ \ K_{\mathrm{mb1}}\approx -0.7,\\
&&B: \ \ \ \ K_{\mathrm{mb1}}\approx -0.3.\nonumber
\end{eqnarray}
One should also perform the summation over higher $s$-, $p$-, and
$d$- intermediate states. We estimate this contribution as
\begin{equation}
\label{KMB3} A,B:  \ \ \ \ K_{\mathrm{mb2}} \approx -0.2
\end{equation}
Combining (\ref{KMB2}) and (\ref{KMB3}) one finds the leading many-body
correction to the enhancement coefficient
\begin{eqnarray}
\label{KMB4}
&&A: \ \ \ \ K_{\mathrm{mb}}\approx -0.9,\\
&&B: \ \ \ \ K_{\mathrm{mb}}\approx -0.5.\nonumber
\end{eqnarray}
It is substantially smaller than the single particle contribution
(\ref{KSP3}), so the many-body perturbation theory is convergent.
The final result for the electron EDM enhancement coefficient,
$K=K_{\mathrm{sp}}+ K_{\mathrm{mb}}$, reads
\begin{eqnarray}
\label{KMB5}
&&A: \ \ \ \ K \approx -6.4,\\
&&B: \ \ \ \ K \approx -3.3.\nonumber
\end{eqnarray}
We recall that the case A corresponds to the energy splitting (\ref{A}),
and the case B corresponds to (\ref{B}).

\section{conclusion}
We have  calculated the electron EDM enhancement coefficient in
the Gd$^{3+}$ ion. The single particle contribution as well as the
leading many-body corrections have been taken into account. The
result is $K\approx -4.9 \pm 1.6$. The main reason for such a large
uncertainty lies in the unknown energy splitting $E_{4f}-E_{5d}$.
Experimental and/or theoretical determination of the
splitting in Gd$^{3+}$ would be the most important step for
improving the accuracy of the enhancement coefficient calculation.

\acknowledgments We are grateful to S. K. Lamoreaux,  L. R.
Hunter, and D. Budker, who attracted our attention to the problem.
We would also like to thank them, as well as W. R. Johnson, for
stimulating discussions and interest in the work.

\end{document}